\documentclass[twocolumn,showpacs,preprintnumbers,amsmath,amssymb]{revtex4}
\input epsf.sty
\usepackage{graphicx}% Include figure files
\usepackage{dcolumn}% Align table columns on decimal point
\usepackage{bm}% bold math

\begin{document}

%\preprint{/123-QED}

\title{Quantum-number projection in the path-integral renormalization group method
}% Force line breaks with \\

\author{Takahiro Mizusaki$^{1,2}$ and Masatoshi Imada$^{2,3}$}
\affiliation{$^1$Institute of Natural Sciences, Senshu University, Higashimita, 
Tama, Kawasaki, Kanagawa, 214-8580, Japan \\
$^2$Institute for Solid State Physics, University of Tokyo,  
Kashiwanoha, Kashiwa, 277-8581, Japan}
\affiliation{
$^3$ PRESTO, Japan Science and Technology Agency
}


\date{\today}% It is always \today, today,

\begin{abstract}
We present a quantum-number projection technique which enables us to exactly treat spin,
momentum and other symmetries embedded in the Hubbard model. 
By combining this projection technique, we extend the path-integral
renormalization group method to improve the efficiency of numerical computations.
By taking numerical calculations for the standard Hubbard model and 
the Hubbard model with next nearest neighbor transfer, we show that
the present extended method can extremely enhance numerical accuracy and 
that it can handle excited states, in addition to the ground state.

\end{abstract}

\maketitle

\section{Introduction}

Quantum many-body systems often possess several symmetries.
For example, the Hubbard model preserves
total spin, total momentum, and some geometrical symmetries on a lattice.
It is crucially important to identify the symmetry and quantum numbers in 
understanding the nature of the ground state, where a symmetry breaking,
for example, often occurs in the thermodynamic limit.  The symmetry
should be restored in finite size systems. However even in finite-size systems, the ground state and excitation spectra reflect
the natures in their thermodynamic limits.
Their excitation spectra and spectroscopic properties are resulted from eigenstates of 
specified quantum numbers and play crucial roles in elucidating the nature of low-energy 
phenomena in condensed matter physics.

To investigate quantum many-body problems, quantum Monte Carlo (QMC) approaches have been one 
of useful methods and  can give ground state properties if there is no minus 
sign problem~\cite{Hatsugai}.
However, they can not fully take an advantage of symmetry explicitly and excitation spectra have not been
well explored.  Furthermore, if the minus sign problem becomes serious as in the case of the
Hubbard model 
on a non-bipartite lattice, QMC methods do not give a well convergent result.
Although the exact diagonalization methods handle the whole excitation spectra, tractable system size is 
severely limited.

In nuclear structure physics, symmetry plays a primarily important role.
For instance, as nucleus is a finite system, rotational symmetry is specially 
important.
Therefore symmetry has been continuously focused in order to solve nuclear quantum 
many-body problems.  
There are several ways to handle symmetries in nuclear structure problems.
Among them, the projection technique is powerful in the respect to broken symmetry 
and its restoration.
In nucleus, a mean field solution is considered as the  first approximation but 
it violates most of symmetries, i.e., total angular momentum, parity, nucleon 
numbers and so on. Then we restore all the symmetries by applying symmetry projection 
(or in other words, quantum-number projection) operators onto symmetry broken mean-field wavefunction. 
Resultant quantum-number projected wavefunction
is known to be able to give a better description. 

Here we consider strongly correlated electrons on a lattice, which have symmetries 
as total spin, total momentum,
and so on. In general, explicit construction of symmetry imposed wavefucntion is quite 
complicated. For instance, wavefunction with a definite total spin needs complicated 
spin coupling among a large number of electrons. 
However, the projection technique enables us to easily handle symmetry imposed 
wavefunction.
This projection method is well harmonized with the recently proposed path-integral 
renormalization group method (PIRG)~\cite{imada1,imada2} which has been quite a powerful tool and is free of 
the notorious minus-sign problem in investigating strongly correlated electron
systems. 
In this method, the ground state is described explicitly by superposition of 
basis states, which often break symmetries possessed by the Hamiltonian
when the numerically manageable number of the basis states, $L$, is limited.
By applying the projection operator to these basis states, we can exactly treat 
the symmetry and extract the state with a specified quantum number.
We show that such a quantum-number projection technique can extensively widen  
applicability  of the PIRG in the following points: (1) Precision of
the numerical calculation is substantially improved. (2) The quantum number of the ground state 
is exactly determined. (3) The extended PIRG by 
the quantum-number projection can handle excited states and spectroscopic properties
in addition to the ground state. 
Such low-energy excitations correspond, in nuclear structure physics, 
to the  {\it yrast} state \cite{Bohr,yrast}, which means
the lowest energy state with specified quantum numbers(for nuclear structure, 
angular momentum).

In Sec. II, we formulate the method of quantum-number projection with examples of spin, spin parity, 
electron momentum and lattice symmetry.  In Sec. III and Sec. IV, we discuss an implementation of the quantum-number 
projection to the algorithm of the path-integral renormalization group (PIRG) method.
Then several different ways of implementation are proposed in the order of increasing elaboration and accuracy.  In Sec. III, we present algorithms and applications from this quantum-number projection technique 
applied afterwards to the obtained PIRG wavefunctions.  Next, in Sec. IV, we show algorithms of the quantum-number projection performed simultaneously with the PIRG procedure, by which the lowest energy state with the specified quantum number is more efficiently extracted. We show that the present methods applied in Sec. III and Sec. IV very efficiently improve the accuracy of the energy estimate.  We show examples  
in the case of the Hubbard model.  We also show how the excitation spectra are obtained.  In the example of the Hubbard model with geometrical frustration effect, the present method enables to 
obtain the ground state as well as excitation spectra, which cannot be obtained in the existing methods.
In Sec. V we summarize the results.

\section{Quantum-Number Projection}
In general, a basis state $\left| \psi  \right\rangle $ described 
by single Slater determinant does not often  satisfy definite symmetry 
properties. Therefore, it can contain many components with unfixed quantum numbers,
 most of which are unnecessary for considering 
the specific eigenstate of considered system.
Here we consider a method to project out a component with a given quantum number 
from such a symmetry broken basis state.

Projection operator ${\cal L}$ is usually defined as 
${\cal L}^2 ={\cal L}$.
If we act ${{\cal L}}$ onto wavefunction $\left| \psi  \right\rangle $, 
${\cal L}\left| \psi  \right\rangle$ contains a component with the considered 
symmetry.
By such quantum-number projected bases, the corresponding projected matrix elements
are evaluated by 
$\left\langle \psi  \right|{\cal L}\left| \psi  \right\rangle $,
$\left\langle \psi  \right|\hat H{\cal L}\left| \psi  \right\rangle $
and $\left\langle \psi  \right|\hat O{\cal L}\left| \psi  \right\rangle $,
for norm,  Hamiltonian and other physical observable matrix elements,
respectively, where $\hat H$ is Hamiltonian and $\hat O$ means a physical observable. 
Note that commutable property between observables and projection operator and 
projection property ${\cal L}^2 ={\cal L}$ simplify
projected matrix elements.
For the physical variables, we assume that $\hat O$ and $\cal L$ are commutable each other.
In this section, we discuss the spin, momentum and lattice symmetries.

\subsection{Spin projection}

Quantum mechanically, finite object with a fixed shape must be rotated to recover the original symmetry.
For nucleus, mean-field methods such as Hartree-Fock and Hartree-Fock-Bogoliubov
approximations, give its optimum 
wavefunction. Though the rotational symmetry in the obtained wavefunction 
is broken, it directly relates the geometrical shape of nucleus.
Restoration of rotational symmetry can be carried out by superposing rotated 
wavefunction. 
This superposition can be exactly carried out by angular momentum projection. 
Rotation in three dimensional space is specified by the Euler's angles and the restoration of the symmetry is usually described by the integration over the Euler's angles and weight of
such superpositions is given by Wigner's $D$ function. Angular momentum projection can be achieved 
by three-fold integration over Euler's angles as we will show later. 
Though this derivation is shown in nuclear structure textbook~\cite{ring}, in Appendix A,
we discuss some properties of the projection operator. 

Here we first consider the spin degrees of freedom of electrons. Though the spin has no relation to any definite shape,
algebraic structure is the same.
As the derivation of angular momentum projection relies on the SU(2) structure,
the same technique can be applied to electron's spin coupling.
We consider to  pick out the total-spin $S$ component from a basis state described by
a Slater determinant. 
As the Slater determinant has a definite number of up and down electrons ($N_\uparrow $ and
$N_\downarrow$), $z$-projection of the spin is $N_0=\frac{N_{\uparrow}-N_\downarrow}{2}$.
This fact simplifies a projection operator to a rather simple one. 
In nuclear structure physics,
it corresponds to the case of angular momentum projection for axially symmetric shape. 

The spin projection operator  has a form as
\begin{equation}
L_{MK}^S\equiv {\frac{2S+1}{8\pi ^2}}\int{d\Omega D_{MK}^{S*}(\Omega )}R(\Omega), \label{eqn:1}
\end{equation}
where $\Omega=(\alpha,\beta,\gamma)$ is Euler angle and 
$D_{MK}^{S}( \Omega)$
is Wigner's $D$ function. Here $M$ and $K$ specify the $z$ component of the total spin, $S_z$.
As explained in the Appendix A, Eq.(\ref{proj_def}), this projection operator operating as
$L_{MK}^S\left| \psi \right.\rangle $ to a state $|\psi \rangle$ filters out $K$ component
of  $|\psi \rangle$ and generate a state which has $S_z=M$ by rotation.
The rotation operator $R( \Omega)$ is defined as 
\begin{equation}
R(\Omega) = e^{i\alpha S_z}e^{i\beta S_y}e^{i\gamma S_z}, \label{eqn:2}
\end{equation}
where $S_y$ and $S_z$ are $y$ and $z$ components of spin operator, respectively.
Wigner's $D$ function is defined by this rotation operator as
\begin{equation}
D_{MK}^S(\Omega)=\langle{SM}|R(\Omega)| {SK} \rangle =e^{i\alpha M}e^{i\gamma K}d_{MK}^S(\beta), 
\label{rot}%{eqn:3}
\end{equation}
where $d_{MK}^S(\beta)=\left\langle{SM} \left|e^{i\beta S_y} \right|{SK}\right\rangle$.
By this projector, the spin projected state is written as
\begin{equation}
L_{MK}^S\left| \psi \right.\rangle =L_{MN_0}^S\left| \psi \right.\rangle,  \label{eqn:4}
\end{equation}
where $N_0=(N_\uparrow-N_\downarrow )/2 $.
Note that $| \psi  \rangle$ has a definite $S_z$ value, $N_0$, but
$e^{i\beta S_y}$ generates different $S_z$ components. Therefore
successive $e^{i\alpha S_z}$ selects finally needed $S_z$ components.
Although the $S_z$ value is not unique and can have values in the range 
$ |S_z| \le S$ in the case of  $S\ne 0$,
this degree of freedom \cite{rme} is eliminated by  the following property of the spin projector; 
\begin{equation}
L_{MK}^SL_{M'K'}^{S'}=L_{MK'}^S\delta _{SS'}\delta _{KM'}. \label{eqn:5}
\end{equation}
This relation can be easily proven by Eq.(\ref{proj_def}) in Appendix A.
This relation shows that spin projection operator satisfies an extended
projection property.
As the PIRG basis states have a definite $z$-component of spin, 
the following relation is satisfied; 
\begin{equation}
L_{N_0M}^S L_{M' N_0}^{S}=L_{N_0,N_0}^S \delta _{MM'}
\label{proj2}
\end{equation}
as the special case of Eq.(\ref{eqn:5}).  Here we note that $L_{N_0,N_0}^S$ has a simpler form, which involves only one-dimensional 
integral, as
\begin{equation}
L_{N_0N_0}^S\equiv {\frac{2S+1}{ 2}}\int_0^\pi {d\beta \sin \beta d_{N_0N_0}^S(\beta )}e^{i\beta S_y}.\label{eqn:7}
\end{equation}

In eq. (\ref{proj2}), we can take $N_0$ as the $M$ value. In this case, as the spin projection operator, 
we can use $L_{N_0N_0}^S$ which satisfies usual projection property 
$\left(L_{N_0N_0}^S\right)^2=L_{N_0N_0}^S$. Therefore in a later discussion, the
spin projection operator is simply denoted as ${\cal L}^S=L_{N_0N_0}^S$ by suppressing $S_z$ value.

Because ${\cal L^S}$ and $H$ commute each other, 
$\langle \psi'|{\cal L^S}H{\cal L^S}|\psi\rangle=
\langle \psi'|H{(\cal L^S)}^2|\psi\rangle=\langle \psi'|H{\cal L^S}|\psi\rangle$ is satisfied. Consequently, norm, Hamiltonian and other physical observable matrix elements between spin-projected basis of $| \psi' \rangle$ and
$| \psi \rangle$  are shown as
\begin{equation}
\left\{ 
\begin{array}{c}
N\\
H\\
O
\end{array}
\right\} = {\frac{2S+1}{2}} \int_0^\pi {d\beta \sin \beta d_{N_0N_0}^S(\beta)\langle {\psi' }|}
\left\{ 
\begin{array}{c}
1\\
\hat H\\
\hat O
\end{array}
\right\}| {\psi (\beta)} \rangle, \label{integ}
\end{equation}
where rotated basis in spin space is defined as 
\begin{equation} 
| {\psi(\beta )}\rangle =e^{i\beta S_y}| \psi\rangle. 
\end{equation}
Here we assume that $\hat O$ is a scalar operator for spin rotation and $S^z$ and $\hat O$ commutes~\cite{rank}.
Note that, $| \psi\rangle $ is a direct product as 
$| \psi \rangle \equiv | \left.{\psi _\uparrow } \right\rangle |\left.{\psi _\downarrow} \right\rangle $,
while its rotated one needs a larger representation space as the up and down components 
are mixed.

For the case that the electron numbers of up and down spins are the same,
the $d$ function simply reduces to Legendre function $P_S (\cos \beta)$, 
\begin{equation}
d_{0,0}^S(\beta)=P_{S}(\cos \beta).
\end{equation}

Involved integral in eq.(\ref{integ}) can now be efficiently evaluated by the Legendre-Gauss
quadrature in practical numerical calculations.  
This quadrature needs less mesh points than those of trapezoidal formula.
Typically, for $S=0$ of the half-filled electron system in 6$\times$6 and 12$\times$12 lattices,
we needs 12 and 24 mesh points, respectively, for numerical convergence.
As spin goes up, larger number of meshes is needed.  

\subsection{Spin-parity projection}

We consider partial spin projection for the restricted case  
that the electron numbers of up and down spins are the same.
Although it is not general, its scope is still wide.

Now we consider the interchange between up and down spin components
and define a parity for this interchange. We show that the parity 
classifies the even and odd total spins. Hereafter we call it spin parity.

The parity operator may be defined as 
$P=\exp(-i \pi S_y) = -i S_y$, where we obtain
\begin{equation}
\langle S0|\exp(-i \pi S_y)|S0 \rangle =d^S_{00}(\pi)=P_S(\cos \pi)=(-)^S.
\end{equation}

This reads that $+$ parity wavefunction corresponds 
to even values for $S$ and $-$ parity wavefunction does to odd values.
Therefore, this spin parity projection 
\begin{equation}  
{\cal L}^{S_{\pm}}=(1 \pm P)/{2}
\end{equation}
yields to the classification between even and odd total-spin states.

The spin-parity projected matrix elements are shown by
\begin{equation}
\left\{ 
\begin{array}{c}
N\\
H\\
O
\end{array}\right\} = \sum_{\sigma =\pm 1 } {(-)^\sigma \left\langle {\phi }
 \right|\left\{ 
\begin{array}{c}
1\\
{\hat H}\\
{\hat O}\\
\end{array}
 \right\}\left| {\phi _\sigma } \right\rangle },
\end{equation}
where $\left| {\phi _\sigma } \right\rangle $ with $\sigma =+1$ and $-1$ takes $\left| \phi  \right\rangle$
and  $P \left| \phi  \right\rangle$, respectively.

If we take the spin projection operator, the spin-parity projection becomes redundant.
However, in the case of multiple quantum-number projection operators, numerical calculations
inevitably become heavy. Since the whole spin projection is much more computer-time consuming, 
the spin-parity projection is an alternative way particularly for the method of simultaneous 
quantum-number projection in each step of PIRG as proposed in Sec. IV.

\subsection{Momentum projection}

In systems with translational invariance, the conservation of momentum holds. 
However, a basis state is not necessarily an eigenstate of the
momentum operator. 
By the projection technique, we restore the translational symmetry.
We define the momentum projection operator as 
\begin{equation}
P^{\vec k}={\frac{1}{\cal N}}\sum_j {e^{i(\vec K -\vec k) \vec R_j}},
\end{equation}
where $\cal N$ is the normalization,
$\vec K$ is the momentum operator and $\vec R_j$ is a shift in a lattice
specified by $j$.
By applying this projection operator, we can calculate projected matrix elements as 
\begin{equation}
\left\{ 
\begin{array}{c}
N\\
H\\
O
\end{array} \right\}={\frac{1}{\cal N}}\sum_{j} {e^{-i{\vec k} \vec R_j}}\left\langle {\phi } 
\right|\left\{
\begin{array}{c}
1\\
\hat H\\
\hat O
\end{array} \right\}\left| {\phi \left( {j} \right)} \right\rangle, 
\label{momentum} 
\end{equation}
where $\left| {\phi \left( {j} \right)} \right\rangle $ is
a shifted wavefunction by the shift $j$.
In an $L_x \times L_y$ lattice, the momentum projection requires  $L_x \times L_y$
larger computation efforts than those of unprojected one.

\subsection{Lattice symmetry projection}

In the Hubbard model on  a two-dimensional lattice, there are several geometrical 
symmetries on a lattice
as $x$-reflection, $y$-reflection and $x$-$y$ interchange symmetries.
Their symmetries  can be classified by parity. By the associated parity 
operator $P$, we can define the corresponding projection operator as 
${\cal L}=\frac{1\pm P}{2} $ similarly to the spin-parity projection.

\section{Quantum-number projection to the PIRG states (PIRG+QP)}

\subsection{Algorithm}

We briefly introduce the path integral renormalization group (PIRG) method,
which has recently been proposed for solving strongly interacting electron systems~\cite{imada1,imada2}.
In general, the ground state $| {\psi _g} \rangle$
can be obtained  by applying the projector
$e^{-\tau H }$ to an arbitrary state $| {\phi _{\rm initial}} \rangle $ 
which is not orthogonal to the true ground state as
\begin{equation}
| {\psi _g} \rangle =\lim_{\tau \to \infty }e^{-\tau H}| {\psi _{\rm initial}} \rangle. 
\label{project}
\end{equation}
In this paper, we consider the standard Hubbard model on a two-dimensional square lattice
defined as
\begin{equation}
H=H_K+\sum_iH_{Ui},
\label{Hubbard}
\end{equation}
where
\begin{equation}
H_K=H_t+H_{t'},
\end{equation}
\begin{equation}
H_t=-\sum\limits_{\langle ij \rangle\sigma } {t\left( {c_{i\sigma }^{\dagger}c_{j\sigma }+{\rm H.c.}} \right)},
\end{equation}
\begin{equation}
H_{t'}=-\sum\limits_{\langle kl \rangle \sigma } {t'\left( {c_{k\sigma }^{\dagger}c_{l\sigma }+{\rm H.c.}} \right)} 
\end{equation}
and
\begin{equation}
H_{Ui}=U{\left( {n_{i\uparrow }-{1 \over 2}} \right)\cdot \left( {n_{i\downarrow }-{1 \over 2}} \right)}.
\end{equation}
Here $i$, $j$ represent lattice points 
and  $c_{i\sigma }^{\dagger}$ ($c_{j\sigma }$) 
is a creation (annihilation) operator of an electron with spin $\sigma$
on the $i$-th site.  The summation over $\langle ij \rangle$ is for the nearest neighbor pairs and that over $\langle kl \rangle$ is for the next-nearest neighbor pairs on the 2D Hubbard model on the square lattice.  We impose the periodic boundary condition.

We decompose $\exp[-\tau H]$ into $\exp[-\tau H] \sim [\exp[-\Delta\tau H_K]\prod_i \exp[-\Delta\tau H_{U_i}]]^{\cal N}$ for small $\Delta\tau$, where $\tau={\cal N}\Delta\tau$.
When we use the Slater determinant as the basis functions, the operation of $\exp[-\Delta\tau H_K]$ to a Slater determinant simply transforms to another single Slater determinant. On the other hand, the operation of $\exp[-\Delta\tau H_{U_i}]$ can be performed by the Stratonovich-Hubbard transformation, where a single Slater determinant is transformed to a linear combination of two Slater determinants. 

One of numerical realizations of Eq.(\ref{project}) is PIRG method~\cite{imada1,imada2}.  After the operation of $\exp[-\tau H]$, the projected wavefunction can be given by an optimal form
composed of $L$ Slater determinants as
\begin{equation}
| {\psi ^{(L)}} \rangle =\sum_{\alpha =1}^L {c_\alpha }| 
{\phi^{(L)}_\alpha } \rangle, 
\end{equation}
where $c_\alpha$'s are amplitudes of $| {\phi^{(L)}_\alpha } \rangle $.
Operation of the ground-state projection  can give optimal $c_\alpha$'s 
and $| {\phi^{(L)}_\alpha } \rangle $'s for a given $L$.
Its detailed algorithm and procedure are found in Ref.~\cite{imada2} .

By a finite number $L$, in most cases, it gives an overestimate of the exact energy 
eigenvalue, since this wavefunction satisfies the variational principle. Therefore, a relation between
energy difference $\delta E$ and energy variance $\Delta E$ may be useful to
extrapolate the energy into the true one.
Here the energy difference is defined as
\begin{equation}
\delta E = \langle {\hat H} \rangle - \langle {\hat H} \rangle_g
\end{equation}
and the energy variance is defined as
\begin{equation}
\Delta E={\frac{\left\langle {\hat H^2} \right\rangle -\left\langle {\hat H} \right\rangle ^2}{\left\langle {\hat H} \right\rangle ^2}}.
\end{equation}
Here, $\langle {\hat H} \rangle_g$ represents the true ground-state energy.
For $\left| {\psi ^{(L)}} \right\rangle$, we evaluate the energy $E^{(L)}$ and energy 
variance $\Delta E^{(L)}$, respectively.

If $\left| {\psi ^{(L)}} \right\rangle$ is a good approximation of the true state,
the energy difference $\delta E^{(L)}$ is proportional to the energy 
variance $\Delta E^{(L)}$.
Therefore extrapolating $E^{(L)}$ into $\Delta E^{(L)}\to 0$ 
by increasing $L$ systematically, we can estimate accurate ground-state energy.
 
Now we consider an implementation of the quantum-number projection to the state
obtained by PIRG. 
The PIRG gives approximated wavefunction  for a given $L$ which is composed of
$L$ linear combinations of  $\left| {\phi^{(L)}_\alpha } \right\rangle $.
One possibility to implement the quantum-number projection is to project out as
\begin{equation}
{\cal L}\left| {\psi ^{(L)}} \right\rangle =
\sum\limits_{\alpha =1}^L {c_\alpha }{\cal L}\left| {\phi^{(L)}_\alpha } \right\rangle, 
\end{equation}
where $\cal L$ is a quantum-number projection operator.
We use the same amplitudes $c_\alpha$'s and the same bases  
$\left| {\phi^{(L)} _\alpha } \right\rangle$'s which the PIRG determines. 
On the other hand,
this amplitude $c_\alpha$'s can be easily reevaluated by diagonalization 
by using quantum-number projected bases, that is, we determine $c_\alpha$'s
by solving the generalized eigenvalue 
problem as
\begin{equation}
H^{\cal L}_{\alpha \beta }\vec x = N^{\cal L}_{\alpha \beta }\vec x, 
\end{equation}
where 
$N^{\cal L}_{\alpha \beta }=\left\langle {\phi _\beta } \right|{\cal L}\left| 
{\phi _\alpha } \right\rangle $, 
$H^{\cal L}_{\alpha \beta }=\left\langle {\phi _\beta } \right|H{\cal L}\left| 
{\phi _\alpha } \right\rangle $.
The latter procedure gives a lower energy eigenvalue.
By adding this procedure for the PIRG basis, we evaluate the projected energies and
energy variances, $E^{L}_{\rm proj}$ and $\Delta E^{L}_{\rm proj}$ for each $L$. We can estimate
accurate energy by extrapolating the projected energy into zero variance.

As a result of the application of this procedure, there appear two new aspects. 
One is that the energy estimate becomes more accurate.
In general, correlation energy comes from dynamical and symmetrical origins.
Original PIRG seeks for better basis states which gain both correlation energies 
in a compromised way.  
On the other hand, by the quantum-number projection operator, correlation energy originated in the 
symmetry is exactly evaluated.
Consequently, the projected energy becomes much lower than the unprojected energy at a
given $L$.
If we use sufficiently large $L$, both values are the same 
and become the exact ground state energy.
In  practical problems, however, we have to use finite number $L$ and exact energy is estimated by
extrapolation. Therefore, at the same $L$, better energy is useful for better estimation 
of the exact energy.

The second point is that it enables the evaluation of excitation spectra.
If we use projection technique, evaluation of excited states with different symmetry 
quantum numbers becomes easier.
The PIRG basis states for $L$ still have components of excitations which most likely
belong to low-lying excited states.
By projecting out the component with different quantum numbers from that of
the desired one, we can evaluate such excited states.
We note the lowest energy state with the specified quantum number (namely, the yrast state)
is obtained.

\subsection{Numerical Results --- Comparison to the exact results --- }
We demonstrate how the method of quantum-number projection procedure applied to the PIRG wavefunction 
works by comparing with the exact results.

\begin{figure}[h]
\begin{picture}(300,200)
    \put(0,0){\epsfxsize 200pt \epsfbox{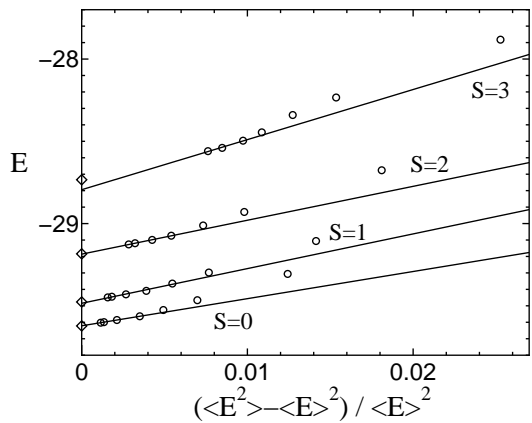}}
\end{picture}
\caption{Extrapolation of the total energy to the zero energy variance for
the spin projection for $S=0,1,2$ and 3 in the 2D Hubbard model with 4 by 4 lattice
and the periodic boundary condition. $L$ is taken up to $L=320$.
The parameters are at $t=1, t'=0$ and $U=4$.
Exact energies with corresponding spin are shown by open diamonds.
}
\label{Spin}
\end{figure}

First we consider the half-filled case 
on $4 \times 4$ lattice with  $U/t=4.0$.
Its exact ground-state energy is -29.62185. The extrapolated energy of the 
PIRG is -29.488, when we use the data up to $L=320$. 
We note that the auxiliary-field quantum Monte Carlo (QMC) method~\cite{Hatsugai} with 
rather large $\tau \sim 20-30$ also gives a similar value to that of the PIRG.  
There is some discrepancy between this energy and the exact one.
This discrepancy comes from the remaining contribution from the higher-spin states contained in the 
projected wavefunction both in the PIRG and the QMC calculations.  
To obtain the real ground-state estimate, we need much larger $\tau$.  
Spin projection can remove it very efficiently. 
In Fig.~\ref{Spin}, we show spin projected energies of $L$=8, 16, 32, 64, 128, 256 and 320 
are plotted as a function of energy variances. The energy variance becomes smaller for larger $L$.
In fact if the correct ground state is given, the variance becomes zero.
As these energies are well scaled linearly as functions of the energy variance when the variance is small,
the extrapolation to the zero variance works well.
The extrapolated ground state energy is -29.611, which is quite close to the exact one.
This result can also be compared with the variational Monte Carlo calculation with the Gutzwiller projection~\cite{YokoyamaShiba}, which gives
-29.47 \cite{4by4mcsinglet}. The SU(2) symmetric Monte Carlo calculation~\cite{Assaad} gives much better estimate~\cite{4by4mcsinglet} with a reasonable value of $\tau\sim 20$. 
This is similar to the PIRG with the spin projection.
In this sense, exact treatment of spin quantum number is crucial in obtaining the exact ground state in an efficient way in the present case.  

In Fig.~\ref{Spin}, projected energies with $S=1\sim3$ are also shown as functions of the energy
variance.
In addition to the ground state, excited states with $S=1\sim3$ have a good linear
scaling. 
Thus we can evaluate energies of the excited states with different spins by the present spin projection 
technique.
This fact shows an essential advantage of the PIRG combined with the quantum-number projection technique, if one compares with the other type of numerical methods including the Monte Carlo methods.

We investigate these extrapolations more closely.
For $S=1$ and $S=2$, extrapolated energies are very close to the exact ones, while
for $S=3$, the extrapolated energy is, to some extent, deviated from the exact one.
As the PIRG is the projection to the ground state, the obtained wavefunction represents the 
ground state approximately. Therefore, as the total spin increases, amplitudes of 
$S\ne0$ components in the PIRG wavefunction are expected to become smaller, 
because such high-energy component is already efficiently eliminated out by the PIRG projection process.
Therefore extrapolated energy for higher-spin (for example $S=3$ state) is worse than those of lower-spin (for example, $S=0$ and $S=1$ states), because the higher energy states are almost missing in the PIRG states.  
Moreover, at a fixed $L$, the variance  becomes larger as the spin goes up, which makes the extrapolation worse.
This also indicates that the quality of projected wavefunctions becomes worse.
We propose an improved algorithm to solve this difficulty for excited states in Sec. IV.

\begin{figure}[h]
\begin{picture}(300,200)
    \put(0,0){\epsfxsize 200pt \epsfbox{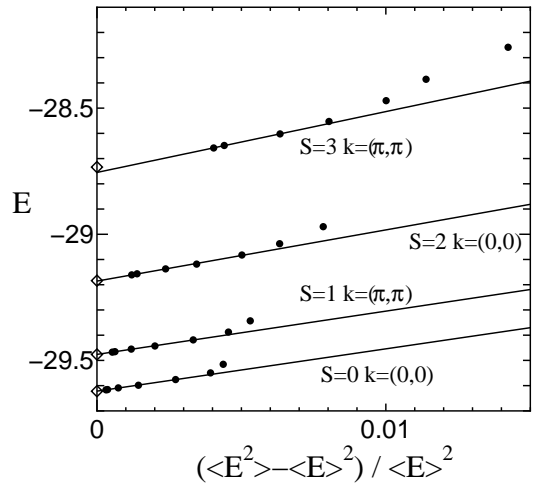}}
\end{picture}
\caption{Extrapolation of the total energy to the zero energy variance for
the spin and momentum projections for ( $S=0,2$ and $\vec k=(0,0)$ )  and 
  ( $S=1,3$ and $\vec k=(\pi,\pi)$ ) in the 2D Hubbard model with 4 by 4 lattice
and the periodic boundary condition. $L$ is taken up to $L=320$.
The parameters are at $t=1, t'=0$ and $U=4$.
Exact energies with the corresponding spin and momentum are shown by open diamonds.
}
\label{SpinMomentum}
\end{figure}

\begin{figure}[h]
\begin{picture}(300,200)
    \put(0,0){\epsfxsize 200pt \epsfbox{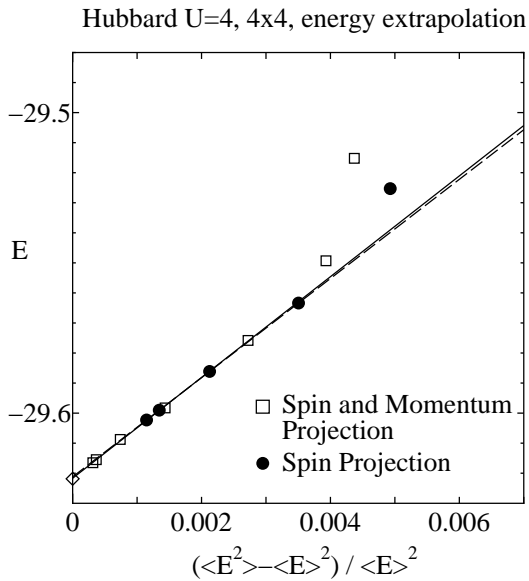}}
\end{picture}
\caption{Detailed comparison of extrapolation of the total energy to the zero energy variance for
the spin projection and spin-momentum projection for $S=0$ ground state 
in the 2D Hubbard model with 4 by 4 lattice
and the periodic boundary condition.
The parameters are at $t=1, t'=0$ and $U=4$.
Exact energy with corresponding spin and momentum is shown by open diamond.
}
\label{SpinandSpinMomentum}
\end{figure}

Next we consider the spin-momentum projection. 
For the even or odd $S$, $\vec k=(0,0)$ or  $\vec k=(\pi,\pi)$ is considered, 
respectively.
In Fig.~\ref{SpinMomentum}, we plot the spin-momentum projected energies as functions of energy 
variances. 
A remarkable difference between spin projection and the spin-momentum projection lies in 
the precision of energy.
The extrapolated energy of the ground state is -29.62166 at $S=0$ and $\vec k=(0,0)$.  The accuracy is one order of magnitude better than the case of the spin projection only.  
As we show in Fig.~\ref{SpinandSpinMomentum}, the spin-momentum projected energy at $L=320$ is -29.61650 while the energy with the spin projection only is -29.60228 for the same $L$. 
With the spin-momentum projection, at the same $L$, the energy becomes lower and extrapolated
energy becomes closer to the exact one than that of the spin projection only.

The higher spin state at $S=3$ with the spin-momentum projection, to some extent, has a better 
extrapolated energies than the spin projection only, while there still remains a tendency that the extrapolation becomes worse
as the total spin goes up or the excitation energy increases. 
To overcome this defect, we have to consider the PIRG with projected bases, namely QP-PIRG
method. We will show the efficiency of QP-PIRG in Sec. IV.

We next study the half-filled system at $6\times6$ lattice with $U/t=4$.
\begin{figure}[h]
\begin{picture}(300,200)
    \put(0,0){\epsfxsize 200pt \epsfbox{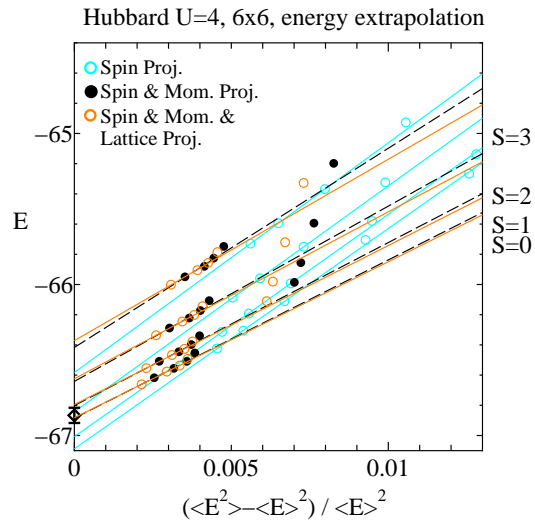}}
\end{picture}
\caption{(color) Detailed comparison of extrapolation of the total energy to the zero energy variance for
the spin projection (blue open circles), spin-momentum projection (filled black circles) and spin-momentum-lattice projection (orange open circles) for $S=0, 1, 2$ and 3 
in the 2D Hubbard model with 6 by 6 lattice
and the periodic boundary condition.
The parameters are at $t=1, t'=0$ and $U=4$.
Quantum Monte Carlo energy for the ground state is shown by open diamond with error bar.}
\label{6by6extrapPIRG+QP}
\end{figure}
In Fig.~\ref{6by6extrapPIRG+QP}, 
we show the extrapolations of spin projected and spin-momentum
and lattice projected energies as functions of the energy variance.
We take the PIRG wavefunctions for various choices of $L$ up to 256.
For the spin projection, we can get the lowest energy states ({\it yrast} states) 
of $S=0, 1 ,2, 3$ from the PIRG wavefunction.
On the other hand, for the spin-momentum-lattice projection, we further resolve
them by their quantum number associated with the corresponding symmetries 
as $S=0,2$ with $ \vec k=(0,0)$ and $S=1,3$ with  $\vec k=(\pi,\pi)$.
Consequently variances of each $L$ wavefunction become smaller.  
Moreover the slopes of the linear extrapolation in the plot of the energy vs. variance
asymptotically obtained at large $L$ for the spin-momentum-lattice projection 
are smaller than the data with the spin projection only.
The spin-only projection shows a slight underestimate of the ground-state energy after the extrapolation,
which is ascribed to an insufficient number of $L$ in this case.
The extrapolated ground-state energy of spin-momentum-lattice projection is -66.8822.
For the sake of comparison, the SU(2) symmetric auxiliary field Monte Carlo 
calculation under the constraint of the spin singlet gives -66.87 $\pm$ 0.05~\cite{4by4mcsinglet}.
Within the statistical error of the quantum Monte Carlo results, 
these two results agree well each other as we see in Fig.~\ref{6by6extrapPIRG+QP}.  
From the extrapolation, the ground state energy is inferred to have better accuracy than the 
Monte Carlo data.

Next we consider the excitation energies. 
The spin projected and the spin-momentum projected approaches give similar values for low-lying states, although the precision is better 
for the latter algorithm.
Spin projected excitation energy of $S=1$ and $S=2$ state is 0.082 and 0.249,
respectively, while the spin-momentum-lattice projection gives 0.081 and 
0.238 for $S=1$, $\vec k=(\pi,\pi)$ and $S=2$, $\vec k=(0,0)$, respectively. 
The accuracy appears to be similar for larger system sizes.

\section{PIRG with Quantum-Number Projected Basis (QP-PIRG)}
\subsection{Algorithm}
In the previous section, we considered the quantum-number projection after the PIRG wavefunction
is obtained for the optimization of the ground state.
To study the properties of excited states, we can further implement an improved algorithm of the quantum-number projection in the PIRG method. That is to perform the PIRG procedure itself
by using the quantum-number projected basis.

In general, the ground-state projector $e^{-\tau H}$ to $\left | \psi \right\rangle$ can 
be applied to lower the energy even within symmetry-imposed restricted space.
When the Hamiltonian preserves some symmetry given by the projection $\cal L$, that is, when $\cal L$ and $H$ are commutable, the lowest-energy state of the specified quantum number, $|\psi \rangle$, can, in principle, be calculated from
\begin{equation}
|\psi \rangle = \mathop {\lim }\limits_{\tau \to \infty }e^{-\tau H}{\cal L}\left| {\psi _{\rm initial}} \right\rangle.
\end{equation}
By introducing the Stratonovich-Hubbard transformation, however, a partial sum over the Stratonovich auxiliary variable destroys the symmetry.  Therefore, if one wishes all the time to keep the symmetry of the state  with the specified quantum number, in an elementary PIRG procedure of the projection $\exp(-\Delta \tau H)|\psi \rangle$, we need to perform the quantum-number projection everytime as ${\cal L} \exp(-\Delta \tau H)|\psi \rangle$ to restore the symmetry.  This is a much more efficient way of obtaining the lowest-energy state with the specified quantum number than the PIRG+QP method discussed in  Sec. III. 

We here explain the algorithm more precisely in the case of the Hubbard model defined by 
Eq.~(\ref{Hubbard}).
The basic procedure is then summarized as repeated operations of $\cal L$ together with the operation of $\exp(-\Delta \tau H)$.  Namely, 
$\mathop {\lim }\limits_{\tau \to \infty }e^{-\tau H}{\cal L}\left| {\psi _{\rm initial}} \right\rangle$
is replaced with $\mathop {\lim }\limits_{\tau \to \infty }[{\cal L} e^{-\Delta\tau H_K}\prod_i{\cal L}e^{-\Delta\tau H_{Ui}}]^{\cal N}\left| {\psi _{\rm initial}} \right\rangle$ by keeping  $\Delta \tau$ small.  Here the operation of $e^{-\Delta\tau H_{Ui}}$ contains the Stratonovich-Hubbard transformation.  A partial and optimized sum of the Stratonovich-Hubbard auxiliary variable constitutes the truncation of basis to keep the number of basis, while it destroys the symmetry.  This algorithm allows the restoration of the required symmetry by the operations of $\cal L$ at 
each step of the truncation.
This is the best way of the optimization to obtain the lowest energy state which has the required symmetry (namely, the yrast states).  
In each step of the operation of $\exp[-\Delta \tau H_K]$ or $\exp[-\Delta \tau H_{U_i}]$, we employ the truncated 
basis which gives the lower energy for the states  ${\cal L}\exp[-\Delta \tau H_K]|\psi\rangle$ or ${\cal L}\exp[-\Delta \tau H_{U_i}]|\psi\rangle$. 
We call this algorithm of simultaneous PIRG and quantum-number projection, Quantum-number Projected PIRG (QP-PIRG).  To differentiate from QP-PIRG, the quantum-number projection procedure using the original PIRG result explained in Sec. III is called PIRG+QP.   

In principle, any quantum-number projection operator can be used in the PIRG. 
However, in practical applications described later, we take a set of multiple projections, namely
spin-parity projection and momentum projection operators, ${\cal L}^{S_\pm} {\cal L}^{\vec k}$.
Ideally, all the quantum-number projection operators should be applied, while it rapidly
increases numerical computation time.
In the present paper, as we study the full momentum dispersion, we employ the momentum projection 
operator. Although the spin projection is important, the spin rotation in spin space mixes
the up and down spin components and we need the twice as large space as the original one for  the Green function in the PIRG procedure, which makes the PIRG computation heavy. Then for the multiple projection of QP-PIRG, to save the computation time, we propose, for a practical use, a combination of the momentum and the spin-parity projection instead of the full spin-momentum projection.
By this approach, the PIRG wavefunction does not have a good spin quantum number. To restore the spin
symmetry perfectly, after the QP-PIRG procedure above, we again
perform the full spin projection afterwards. Namely, to obtain  a final 
result, ${\cal L}^S {\cal L}^{\vec k}{\cal L}_{lattice}$ is applied.  This constitutes the full procedure of
QP-PIRG.

At each quantum-number projection, the integrations or summation such as those in Eqs.(\ref{integ}) and 
(\ref{momentum}), can be very efficiently parallelized in actual computations if parallel 
processors are available.  In each process, we store the Green function $G_{ij}^{(\psi_{\alpha}, {\cal L} \psi_{\beta})} \equiv \langle \psi_{\alpha}|c_i^{\dagger}c_j{\cal L} |\psi_{\beta}\rangle$, while the update of the Green function after the operation of each $ e^{-\Delta\tau H_{Ui}}$ is written as 
\begin{equation}
G_{ij}^{(\psi_{\alpha}, {\cal L} \psi'_{\beta})} \equiv \langle \psi_{\alpha}|c_i^{\dagger}c_j{\cal L} |\psi'_{\beta}\rangle,
\end{equation}
or 
\begin{equation}
G_{ij}^{(\psi'_{\alpha}, {\cal L} \psi_{\beta})} \equiv \langle \psi'_{\alpha}|c_i^{\dagger}c_j{\cal L} |\psi_{\beta}\rangle,
\end{equation}
where 
\begin{equation}
|\psi'_{\beta}\rangle=\sum_{\sigma} \frac{1}{2}\exp[2a\sigma(n_{i\uparrow}-n_{i\downarrow})-\Delta \tau U/2]|\psi_{\beta}\rangle,
\end{equation}
with $\sigma$ being the Stratonovich auxiliary variable and $a=\tanh^{-1} \sqrt{\tanh(\frac{\Delta \tau U}{4})}$.  When one term of the sum over $\sigma$ is taken in the truncation process, the updated Green function is efficiently calculated from the old Green function $G_{ij}^{(\psi_{\alpha}, {\cal L} \psi_{\beta})}$ in the same way as Eq.(3.10)-(3.14) in Ref.~\cite{imada2}.

\subsection{Numerical Results of QP-PIRG}

\subsubsection{Results for 6 by 6 lattice}

\begin{figure}[h]
\begin{picture}(300,200)
    \put(0,0){\epsfxsize 200pt \epsfbox{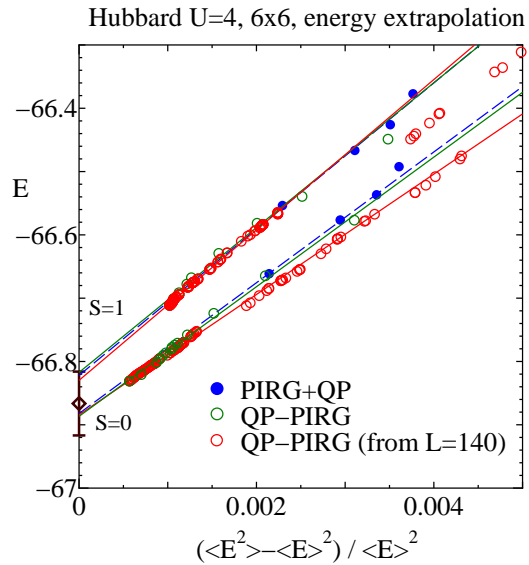}}
\end{picture}
\caption{(color) Extrapolations of the energy to the zero energy variance by using 
PIRG+QP (filled blue circles and dotted lines) and QP-PIRG (open green circles and solid lines) for the 2D Hubbard model with 6 by 6 lattice
and the periodic boundary condition.
The parameters are at $t=1, t'=0$ and $U=4$.
The ground-state energy of Monte Carlo calculation is also shown by open brown diamond at zero variance with
the statistical error bar (-66.8664 $\pm$ 0.0504). 
The red symbols with red solid lines are derived from the largest $L$ wavefunction ($L=140$) of QP-PIRG, where we 
choose partial $L_a$ basis functions which have the largest weights among $L$ bases.  The plots are 
obtained with increasing $L_a$ up to $L=140$.
}
\label{6by6extrap}
\end{figure}

Now we show numerical results of QP-PIRG. 
We first show the case of 6 by 6 lattice at $U=4$ and $t=1, t'=0$.  
In Fig.~\ref{6by6extrap}, we show the extrapolation of QP-PIRG result 
by green open circles
by using the projection up to $L=140$. 
As we discussed, the QP-PIRG with quantum-number projected bases seeks 
for optimum {\it yrast} states concerning the considered symmetry in every PIRG process.
In this calculation, we took spin-parity and momentum projection operators.
For $S=0$ and $\vec k=(0,0)$ state, we use ${\cal L}^{S_+}{\cal L}^{\vec k=(0,0)}$.
As the obtained wavefunction still contains $S=2,4,..$ components,
we apply ${\cal L}^{S=0}{\cal L}^{\vec k=(0,0)}{\cal L}_{lattice}$ 
projection operators afterwards for final results.  

This QP-PIRG can generate 
a better wavefunction than the PIRG+QP state as we see in the comparison with blue closed circles. 
Here we show results of the PIRG+QP state obtained after spin-momentum projection. 
In fact, for $S=0$ with $\vec k=(0,0)$ state,
in the PIRG+QP result even at 
$L=256$, the energy is -66.5765, while the same energy can be given 
at $L\sim15$ by the QP-PIRG. 
This means that for the ground state,  
basis states are more elaborately selected by the QP-PIRG. 
Thus, the quantum-number projection simultaneously with the PIRG provides an efficient way of obtaining better wavefunctions.   
The extrapolated ground-state energy is -66.879 which is well within the 
statistical error of the previously cited Monte Carlo energy.  In fact, from the extrapolation procedure in Fig.~\ref{6by6extrap}, the accuracy of the QP-PIRG seems to have more than 4 digits and is higher than the accuracy of the presently referred quantum Monte Carlo result~\cite{4by4mcsinglet}, since the energy at $L=140$ is already lower than the upper bound of the Monte Carlo estimate. 
Namely, the QP-PIRG result seems to give the 
highest accuracy among these comparisons.  

In addition, we have also shown in Fig.~\ref{6by6extrap} an alternative way of the extrapolation.   
The red symbols are derived from the largest $L$ state after QP-PIRG, where $L=140$ in this case.  This state is represented by $L$ basis functions.  After ordering these basis functions from the largest weight in the linear combination,  we may truncate the basis functions by taking only the $L_a$ states from that with the largest weight.  By using these truncated functions with different $L_a$, we have plotted the energy and variance of these truncated states.  This gives very close estimate to the QP-PIRG result shown above as the open green circles. 
A small difference between this procedure and the original QP-PIRG is seen at larger variance. This may be due to the fact that at small $L$, the present truncation at small $L_a$ does not necessarily give the lowest energy state with $L_a$.  Another possible origin is that the iteration of the present QP-PIRG is not sufficient in reaching the lowest energy state under the constraint of each $L$. In any case, the linearity of the plot in the plane of the energy and the variance is well satisfied in both cases, particularly for the latter procedure,  and the asymptotic slopes at large $L$ look the same. 

\subsubsection{Results with next-nearest neighbor transfer}

In the previous section, we consider the standard Hubbard model with $t'=0$.
Conventional quantum Monte Carlo calculation could be
performed to investigate such ground state properties.
The PIRG is an alternative method in this respect while 
it and its extension have an advantage in investigating
the excitation spectra. Especially, quantum-number projection enables us to handle 
{\it yrast} states with the same effort as the ground state.
However, it is expected that the 2D Hubbard model with $t'=0$ has an 
antiferromagnetic long-ranged order in the thermodynamic limit and has a simple low-energy
structure.  To test the efficiency of our algorithm in a more severe condition,
we investigate the extended Hubbard model 
by including the next-nearest neighbor transfer, which causes the geometrical 
frustration effect.  The quantum Monte Carlo method is known to have a severe difficulty
when $t'$ becomes large.  

Recently  by using the PIRG method, 
the non-magnetic insulator (NMI) phase has been found near the Mott transition 
for relatively large $t'$~\cite{imada3}. This phase can not be investigated by the Monte Carlo methods due to 
severe minus sign problems.
Therefore, the PIRG is so far the only technique suited for this study.
Here we explore how the present quantum-number projection technique 
improves the precision of the PIRG in such a study.
Here we consider the half-filled system on 4 by 4 lattice with $U/t=5.7$ and 
$t'=0.5$.
Monte Carlo method does not give us convergent results because of the minus sign problem at this parameter value. We compare our results
with the exact one. 

\begin{figure}[htb]
\begin{picture}(300,200)
    \put(0,0){\epsfxsize 180pt \epsfbox{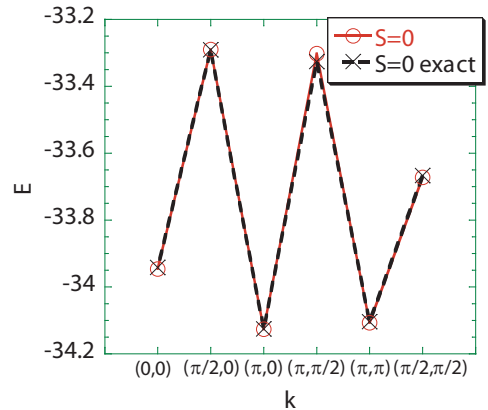}}
\end{picture}
\caption{(color) The energy dispersion of the 2D half-filled Hubbard model at $U=5.7, t=1, t'=0.5$
for $S=0$ states. The system size is 4 by 4 with the periodic boundary
condition. The comparison with the exact results (black crosses) shows that the QP-PIRG (red circles) works excellently well for the ground state as well as the dispersion even when the 
geometrical frustration effect is large. 
}
\label{dispersion1}
\end{figure}
\begin{figure}[h]
\begin{picture}(300,200)
    \put(0,0){\epsfxsize 180pt \epsfbox{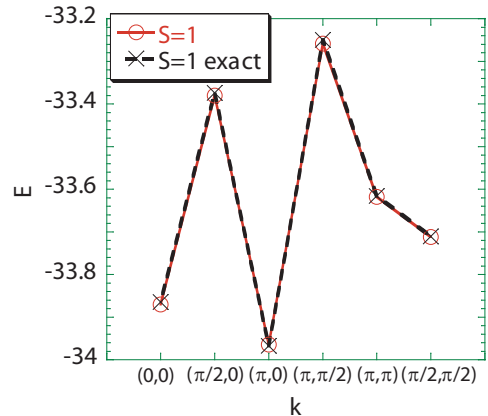}}
\end{picture}
\caption{(color) The energy dispersion of the 2D half-filled Hubbard model at $U=5.7, t=1, t'=0.5$
for $S=1$ states. The system size is 4 by 4 with the periodic boundary
condition.  The comparison shows that the QP-PIRG (red crosses) works excellently well even for the spin excitations.
}
\label{dispersion2}
\end{figure}

In Figs.~\ref{dispersion1} and \ref{dispersion2}, we show comparison of the dispersions obtained by the QP-PIRG 
with the exact diagonalization result.
This system has the ground state at $S=0$ and $\vec k = (\pi,0)$.
The $S=0$ with $\vec k = (\pi,\pi)$ state severely competes with this ground state. 
The lowest-energy $S=1$ state has $\vec k=(\pi,0)$. This energy is very close to those of
doubly degenerate $\vec k=(0,0)$ states and the second lowest state with $S=1$ and 
$\vec k=(\pi,0)$.
The comparison of QP-PIRG (red circles) and the exact diagonalization results (black crosses) in Figs.~\ref{dispersion1} and \ref{dispersion2}
indicates excellent agreement. In general, the errors are less than 0.01, which means the accuracy higher than the 4 digits. 

\section{Summary}

We have presented the quantum-number projection technique and its implementation to
the PIRG method, both of which works well irrespective of the details of the considered system.
The quantum-number projection method can pick up a component with required symmetries 
from symmetry broken wavefunctions (i.e., mean field wavefunction and so on).
In the Hubbard-type model, the symmetries have a significant role in the low-energy states. 
In particular, spin, momentum and lattice symmetries play specially important roles
in determining the low-energy states.
Restoration of the spin symmetry can be carried out by taking a spin
projection operator, which is the same technique as the angular momentum 
projection in nuclear structure physics.
Spin rotation is performed in the spin space and 
the spin projection is represented  by one dimensional integral for the rotation.
The momentum projection is simply given from the superposition of spatially translated 
basis functions.  We have also considered geometrical
symmetry on a lattice for projections such as the inversion and rotation symmetries. 

Quantum-number projection operator $\cal L$ is represented by the sum of exponential of one-body operator.
In the PIRG, the wavefunction is expressed by a linear combination of $L$ basis states, while the symmetries are not retained in each basis state in general.
Then the quantum-number projection is efficiently introduced for each basis state.
In the present paper, we have introduced two ways of implementing quantum-number projection 
into the PIRG. One way is to carry out quantum-number projection afterwards for the already obtained PIRG wavefunction (PIRG+QP).
The ground state is efficiently extracted by specifying the quantum number with higher accuracy than the 
PIRG only.  Although the PIRG does not efficiently pick up the excited states, we can obtain
several low-lying excited states with various symmetries from the PIRG wavefucntion, if a small portion of the excited states still remain after the PIRG procedure.
Other is to carry out the PIRG by using quantum-number projected basis states (QP-PIRG).
By this extended PIRG, we can precisely evaluate excitation spectra.  Although QP-PIRG requires more computation time, the accuracy of the ground state is more
improved than PIRG+QP, particularly for the excitation spectra.

In numerical calculations, quantum-number projection can be performed exactly.
Moreover, as quantum-number projection operators $\cal L$ are commutable with Hamiltonian $H$,
the relation ${\cal L}H{\cal L}= H{\cal L}$ simplifies numerical calculations.
As examples, the accuracy and efficiency of the algorithm 
has been tested for the standard Hubbard model on two-dimensional square lattice 
as well as for the 2D Hubbard model with nonzero next nearest neighbor  
transfer, where geometrical frustration effects are large. 
We have shown that the quantum-number projection implemented to the PIRG excellently works.
More concretely, the spin projection and spin-momentum projection by PIRG+QP greatly
improve the accuracy of energy. QP-PIRG further improves the accuracy of the extrapolated energy.  
This algorithm also enables accurate calculations of low-lying excitation spectra with 
different quantum numbers from those of the ground state.  The energy dispersions of the specified 
total spin have been shown to give highly accurate results, particularly by using the QP-PIRG method.
This accuracy does not depend on the details of the lattice structure and the dimensionality.
In our examples the accuracy becomes higher or comparable to 4 digits.

When the system size increases in the 2D Hubbard model, we do not have a relevant clue to judge the 
accuracy of the calculation by the present algorithm.  On the half-filled case, however, we can compare 
the results with the quantum Monte Carlo results and the agreement is satisfactory. 
To reach the same accuracy, it seems to be necessary to increase the number of the basis functions $L$ gradually with the increase of the system size.  

\section*{ACKNOWLEDGEMENTS}
We are grateful to F.F. Assaad for useful discussions and providing the data cited in Ref.~\cite{4by4mcsinglet}. A part of computation has been performed by using the facilities of Supercomputer Center, Institute for Solid State Physics, University of Tokyo. 
\appendix
\section{}

In this appendix, we discuss some properties of the spin projection
operator.

We expand $|\psi\rangle$ by complete set $|{SM\alpha}\rangle$
regarding to spin quantum number, as
\begin{equation}
| \psi \rangle = \sum_{SM\alpha }c_{SM\alpha}{| {SM\alpha } \rangle },
\end{equation}
where  $ c_{SM\alpha}= \langle{SM\alpha} | 
{\psi}\rangle $
and $\alpha$ denotes other quantum numbers.
Operation of rotational operator $R(\Omega)$
 to  $|\psi\rangle$  results in
\begin{eqnarray}
R(\Omega)\left| \psi\right.\rangle & = & \sum_{SM\alpha } {c_{SM\alpha }R(\Omega)\left|{SM\alpha} \right.\rangle} \nonumber \\
& = & \sum_{SKM\alpha }{c_{SM\alpha }D_{KM}^S(\Omega)\left| {SK\alpha } \right.\rangle }, \label{eqn:27}
\end{eqnarray}
where we use eq.(\ref{rot}). 
By this relation, projection onto $|\psi\rangle$   is represented by
\begin{eqnarray}
L_{MK}^S| \psi\rangle & = & {\frac{2S+1}{8\pi ^2}}\int{d\Omega D_{MK}^{S*}(\Omega)}R(\Omega)| \psi\rangle \nonumber \\
& = & \sum_\alpha{| {SM\alpha } \rangle }\langle {{SK\alpha }} |
 {\psi } \rangle \label{proj_formula},
\end{eqnarray}
where we use the following relation as
\begin{equation}
\int {d\Omega D_{MK}^{S*}(\Omega)D_{M'K'}^{S'}(\Omega)=}{\frac{8\pi ^2}{2S+1}}\delta _{SS'}\delta _{MM'}\delta _{KK'}.
\end{equation}
Therefore, $L_{MK}^S$ projects out $| {SM}\rangle $ component
from $| \psi\rangle $.
By Eq.(\ref{proj_formula}), projection operator is represented as
\begin{equation}
L_{MK}^S=\sum_\alpha{| {SM\alpha } \rangle }\langle {SK\alpha } |.
\label{proj_def}
\end{equation}


\begin{thebibliography}{99}
\bibitem{rme} In nuclear physics, reduced matrix elements are used.
\bibitem{rank}For a tensor operator, several factors are needed.
\end{thebibliography}
\end{document}